\documentclass[]{emulateapj}

\def\Msun{{\rm M}_\odot}
\def\Mstel{M_\ast}

\def\Mstar{M^*}
\def\Mhalo{M_{\rm halo}}

\def\resp{respectively}
\def\bfr{\bf\color{red}}

\def\ssfr{{\rm sSFR}}
\def\sfr{{\rm SFR}}
\def\MS{SFMS}
\def\sigms{\sigma_{\rm \MS}}
\def\phitot{$\Phi$}
\def\phisf{$\Phi_{\rm SF}$}
\def\phiq{$\Phi_{\rm Q}$}
\def\sfrd{{\rm SFRD}(t)}

\def\beq{\begin{equation}}
\def\eeq{\end{equation}}
\def\bitem{\begin{itemize}}
\def\eitem{\end{itemize}}
\def\benum{\begin{enumerate}}
\def\eenum{\end{enumerate}}

\mathchardef\mhyphen="2D

\def\cite{{\bfr CITE}}
\usepackage{graphicx}
\usepackage[bottom]{footmisc}
\usepackage{color}
\usepackage[breaklinks=true]{hyperref}
\usepackage{helvet}
\shortauthors{ABRAMSON ET AL.}
\shorttitle{SMF EVOLUTION FROM LOG-NORMAL SF HISTORIES}

\begin{document}

\title{
Matching the Evolution of the Stellar Mass Function Using\\
Log-normal Star Formation Histories
}

\slugcomment{Submitted to ApJ Letters}

\author{
Louis E. Abramson\altaffilmark{1,$\ast$},
Michael D. Gladders\altaffilmark{1},
Alan Dressler\altaffilmark{2}, 
Augustus Oemler, Jr\altaffilmark{2},
Bianca Poggianti\altaffilmark{3},\\
and Benedetta Vulcani\altaffilmark{4}
}


\begin{abstract}

We show that a model consisting of individual, log-normal star formation histories for a volume-limited sample of $z\approx0$ galaxies reproduces the evolution of the total and quiescent stellar mass functions at $z\lesssim2.5$ and stellar masses $\Mstel\geq10^{10}\,\Msun$. This model has previously been shown to reproduce the star formation rate/stellar mass relation (SFR--$\Mstel$) over the same interval, is fully consistent with the observed evolution of the cosmic SFR density at $z\leq8$, and entails no explicit ``quenching'' prescription. We interpret these results/features in the context of other models demonstrating a similar ability to reproduce the evolution of (1) the cosmic SFR density, (2) the total/quiescent stellar mass functions, and (3) the $\sfr$--$\Mstel$ relation, proposing that the key difference between modeling approaches is the extent to which they stress/address diversity in the (starforming) galaxy population. Finally, we suggest that observations revealing the timescale associated with dispersion in $\sfr(\Mstel)$ will help establish which models are the most relevant to galaxy evolution.

\end{abstract}


\altaffiltext{1}{
Department of Astronomy \& Astrophysics and Kavli Institute for Cosmological Physics, The University of Chicago, 5640 South Ellis Avenue, Chicago, IL 60637, USA
}
\altaffiltext{2}{
The Observatories of the Carnegie Institution for Science, 813 Santa Barbara Street, Pasadena, CA 91101, USA
}
\altaffiltext{3}{
INAF-Osservatorio Astronomico di Padova, Vicolo Osservatorio 5, 35122 Padova, Italy
}
\altaffiltext{4}{
Kavli Institute for the Physics and Mathematics of the Universe (WPI), Todai Institutes for Advanced Study, University of Tokyo, Kashiwa 277-8582, Japan
}
\altaffiltext{$\ast$}{
\href{mailto:labramson@uchicago.edu}{\tt labramson@uchicago.edu}
}


\section{Introduction}
\label{sec:intro}

Three measurements have emerged as central to the study of galaxy star formation histories (SFHs): 
\benum
	\item[1)] the cosmic star formation rate density \citep[SFRD; see][and references therein]{MadauDickinson14}; 
	\item [2)]the stellar mass function, $\Phi(\Mstel)$, for starforming (\phisf) and quiescent (\phiq) galaxies \citep[e.g.,][]{Tomczak14}; 
	\item[3)] the SFR--$\Mstel$ relation, or SF ``Main Sequence" \citep[\MS; e.g., ][]{Brinchmann04,Noeske07,Salmon14}. 
\eenum	
Any viable theory of galaxy evolution must reproduce these observations.

These ``pillars" rest on each other physically:
\beq
	\sfrd=\int\Phi_{\rm SF}(\Mstel,t)\langle\sfr(\Mstel,t)\rangle\,d\Mstel.
\label{eq:integral}
\eeq
Cosmic SFRD evolution depends explicitly on the evolution of the (starforming) stellar mass function and the \MS\ (i.e., $\langle\sfr(\Mstel,t)\rangle$). Depending on the modeling approach, one of these phenomena can emerge naturally if the other two are reproduced, but all serve as checks/constraints regardless of technique. That is, even if a model describes a plausible evolutionary scenario for \phitot\ based on the \MS, it is not an accurate scenario unless the correct $\sfrd$ emerges as a consequence.

In \citet[][hereafter G13]{Gladders13b}, noting that cosmic $\sfrd$ is remarkably log-normal in shape, we posited that this {\it analytic form} might also describe the SFHs of individual galaxies. We developed a model comprised of log-normal SFHs for a volume-complete sample of $z\approx0$ galaxies ($\Mstel\geq10^{10}\,\Msun$, any $\ssfr$) by jointly fitting their observed $\Mstel,\, \sfr$ values and the cosmic SFRD at $z\leq8$. While $\sfrd$ was thus matched by construction, we showed that the model also reproduced observed $\ssfr$ distributions at $0.2\leq z\leq1$, and the $z\sim2$ \MS\ if those distributions were used as constraints. We did not, however, examine the model's implications for the stellar mass function.

Here, we complete the analysis by comparing G13 model predictions to the observed evolution of \phitot\ and \phiq, showing that the data are reproduced with remarkable fidelity. By doing so, we demonstrate that a purely ``galaxy-up" approach -- which addresses diversity in the galaxy population -- can match key ensemble metrics of galaxy evolution previously modeled using mean, statistical relations \citep[e.g.,][]{PengLilly10,Behroozi13}. We close by describing ways to elucidate how well the SFHs of real galaxies fit into these paradigms.


\begin{figure*}[t!]
\centering
\includegraphics[width = \linewidth]{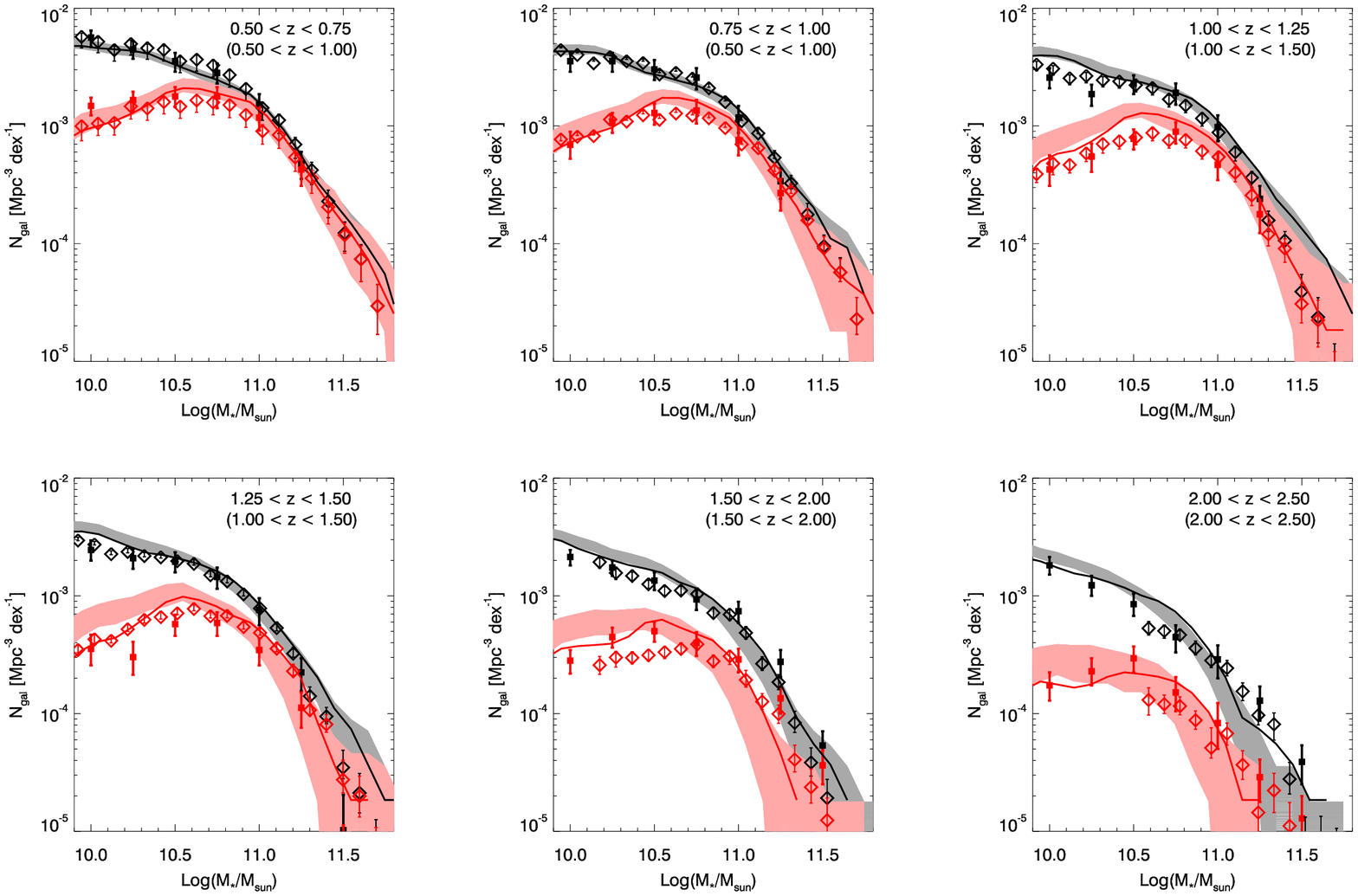}
\caption{The evolution of \phitot\ and \phiq\ since $z\sim2.5$. Diamonds/squares show \citet{Muzzin13} and \citet{Tomczak14} data, \resp. Redshift binning matches the Tomczak et al.\ data; parentheses denote the coarser Muzzin et al.\ intervals from which those data were drawn. Shaded pink/grey bands are \phitot/\phiq, \resp, predicted using the ensemble of log-normal SFHs from \citet{Gladders13b}. Solid lines show results if mergers are neglected. Though no $\Mstel$ information was used to constrain the models, the predicted evolution of the mass functions agrees remarkably well with the data, suggesting the lognormal SFH is consistent with real galaxy SFHs at the mass-bin level, and also that smooth, continuous SFHs can reproduce the observed evolution of the ``quenched'' population.}
\label{fig:SMF}
\end{figure*}

\section{Log-normal Star Formation Histories}

A detailed account of the extraction of SFHs can be found in G13, but we review some critical aspects below. 

\subsection{Modeling}
\label{sec:model}

The fitting procedure mentioned above found the best log-normal SFH parameters (Equation \ref{eq:log-normal}) for 2094 galaxies at $\langle z\rangle\approx0.07$ based on how well the ensemble of such histories matched the observed SFRD$(z\leq8)$. Explicitly, the shape of the SFH of the $i$-th galaxy in a comoving volume $V$ -- drawn from the Sloan Digital Sky Survey \citep[][]{York00} and the Padova Millennium Galaxy and Group Catalogue \citep{Calvi11} -- was: 
\beq
	\sfr(t)_{i}\propto
	\frac{1}{\sqrt{2\pi\tau_{i}^{2}}}
	\frac{\exp\left[-\frac{1}{2}\frac{(\ln t-T_{0,i})^2}{\tau_{i}^2}\right]}{t},
\label{eq:log-normal}
\eeq
where $(T_{0},\tau)$ are the peak-time and width of each SFH in $\ln(t)$, \resp. The fitting determined the best values for these parameters such that (1) individual SFHs -- normalized to the appropriate $\Mstel$ -- produced a galaxy's observed SFR at $z=z_{\rm obs}$, and (2) the ensemble of SFHs summed to the observed $\sfrd$:
\beq
	\sum_{i=1}^{N_{\rm gals}}\sfr(t)_{i}=V\cdot\sfrd,
\eeq
for all $t$ spanned by the SFRD data \citep{Cucciati12}.
 
This approach is extremely flexible, entails no assumptions about the data, and is highly predictive in principle: any SFH-related observable (e.g., colors) could be derived for any redshift spanned by the SFRD data. Of course, these virtues require imposing a form to $\sfr(t)$.

For our final G13 model -- upon which this paper is based -- we included the {\it zeropoint} of the \MS\ and its {\it dispersion} collapsed across $\Mstel$ as additional constraints at various $z\lesssim1$. Thus, the normalization and spread of the SFHs were constrained at $z\lesssim1$, but not $\sfr(\Mstel)$ itself; i.e., no $\Mstel\mapsto\sfr$ mapping was imposed anywhere except implicitly through the $z\approx0$ input data. Likewise, the $\Mstel$ distribution (i.e., \phitot) was set only implicitly by these data (complete to $\Mstel=10^{10}\,\Msun$); it was never used as an explicit fitting constraint.

\subsection{Quiescent Fractions}
\label{sec:pasfrac}

At any epoch constrained by a sSFR distribution, the G13 procedure ensured only that the total fraction of quiescent SFHs matched the total observed fraction of quiescent galaxies. ``Quiescent'' was defined as $\log\ssfr(\Mstel,t) <[\langle\log\ssfr(\Mstel,t)\rangle - 0.6\, {\rm dex}]$; i.e., $\sim1.5\sigma$ below the mean \MS\ relation. Model SFHs could take any value below this threshold.

While the quiescent {\it definition} was thus mass-dependent, the fitter considered only the {\it total number of SFHs meeting it}, so the resulting model $\Mstel$ distributions were unconstrained. This approach affects the normalization, but not the {\it shape} of \phiq. No aspect of our modeling guarantees that the detailed evolution of \phitot\ or \phiq\ will be reproduced, so comparisons are valid tests of our log-normal SFH parametrization.


\section{Results}
\label{sec:results}

\subsection{Predictions for the Evolution of \phitot\ and \phiq}
\label{sec:main_results}

We plot \phitot\ and \phiq\ as measured by \citet{Tomczak14} and \citet{Muzzin13} across $0.5\lesssim z\lesssim2.5$ and overlay our model predictions in Figure \ref{fig:SMF}. Model uncertainties derive from marginalizing over galaxy merger histories (see G13 for details), locations within a redshift bin, and random $\Mstel$ errors of $0.1$ dex \citep[consistent with typical values; e.g.,][]{Kauffmann03}. Solid lines show results if mergers are neglected. 

Two points are key. The first is that the model reproduces the evolution of the shape {\it and} normalization of \phitot\ very accurately, especially when considering uncertainties in the data and merger prescription. Though ``informed'' only of the {\it total} $\dot\Mstel$ evolution of the universe (via $\sfrd$), this result shows that the model {\it put that mass in the correct place} ($\Mstel$-bin) {\it at the correct time}. This suggests that the log-normal parametrization is in fact a good approximation of real galaxy SFHs, at least at the mass-bin level.  

The second point is that the model reproduces the shape of \phiq, especially the turnover at $\Mstel<\Mstar_{\rm Q}$ not exhibited by \phitot\ or \phisf\ \citep[e.g.,][]{Ball06, Moustakas13,Muzzin13,Tomczak14}. Hence, even though the model contains {\it no} explicit quenching prescription, it produces the correct evolution from the starforming to quiescent populations (defined as a cut in $\ssfr$; see Section \ref{sec:pasfrac}) as a function of $\Mstel$ and time. This suggests that the model SFHs for quiescent galaxies are both accurate absolutely -- for a given mass-bin -- and appropriately divergent from those of equal-mass starforming contemporaries (see Figure \ref{fig:SFH}). However, as discussed in Section \ref{sec:quenching}, ``divergent'' in this context does not mean ``subject to additional physics''.


No data from the comparison observations was used at any point except to normalize the total mass functions (grey curves to black points) in the first two panels (top-left). All other panels show epochs in which the G13 model was constrained only by $\sfrd$. Mathematically, this is to say the model was sensitive only to the {\it integral} of \phiq\ and \phisf\ ($=\Phi - \Phi_{\rm Q}$; see Equation \ref{eq:integral}). Thus, the fact that the mass {\it distributions} resemble the data at better than the factor of 2 level at all $\Mstel$ is rather remarkable. Indeed, this is true even at $z\lesssim1$ (except for the normalization of \phitot): no mass-based constraints were employed at these redshifts (Sections \ref{sec:model}, \ref{sec:pasfrac}). 

\subsection{More on the Absence of Quenching}
\label{sec:quenching}

 As just mentioned, our model incorporates no explicit ``quenching''; i.e., SFH discontinuities turning starforming galaxies into non-starforming ones. Because all model SFHs are continuous and smooth, the quiescent label is purely semantic; it is {\it not} an indication that a galaxy belongs to a qualitatively different population. As shown in Figure \ref{fig:SFH}, ``quiescent'' in our model really means ``finished'': the SFHs of these galaxies are distinguished from their starforming peers' only by smaller $(T_{0},\tau)$ values, reflecting earlier, more rapid growth.

That said, our model does not rule-out abrupt quenching events/scenarios, or fundamental bimodality in the galaxy population. It simply suggests that these phenomena are not necessary to reproduce the \MS, \phitot/\phiq, and the evolution of cosmic SFRD.

\subsection{Summary}
\label{sec:results_summary}

At a minimum, the results above combined with those from G13 suggest that the log-normal parametrization is consistent with real galaxy SFHs to at least the level at which the \MS, \phitot\ or \phiq, and \sfrd\ are sensitive to them. However, because they show that these ensemble metrics -- often modeled using the mean behavior of statistical quantities \citep[e.g., $\langle\ssfr(\Mstel,t)\rangle$;][]{PengLilly10} -- can be reproduced using evolutionary tracks for individual galaxies, these results serve more broadly as an endorsement of ``bottom-up'' approaches to galaxy evolution. We explore these issues in the next section. 

\section{Context and Implications}
\label{sec:implications}

We have shown -- here and in G13 -- that a model composed of nothing but log-normal SFHs for individual galaxies is capable of matching three of the most important ensemble observables related to galaxy evolution: 
\benum
	\item[1)] cosmic SFRD (G13; by construction);
	\item[2)] the SFMS at $z\lesssim2$ (G13);
	\item[3)] the evolution of \phitot\ and \phiq\ at $z\lesssim2.5$ (this paper).
\eenum

We are not alone in this accomplishment, however. Others have approached $\sfrd$, \phitot, the \MS, and SFHs in a semi-empirical, (quasi-)holistic fashion \citep[e.g.,][]{PengLilly10, Moster13, Behroozi13, Kelson14, Lu14}. Not all were principally concerned with extracting SFHs \citep[e.g.,][]{PengLilly10}, and not all attempted to match each of these ``pillars'' \citep[e.g.,][]{Lu14}, but the results presented above and in these other works suggest that there are now many ``good'' descriptions of the data. So, how are we to proceed? Have we identified an especially meaningful SFH form, or are the benchmark metrics simply easy to reproduce? How can we determine which is the case?

\begin{figure*}[t!]
\centering
\includegraphics[width = 0.85\linewidth, trim = 0cm 0.25cm 0cm 0cm]{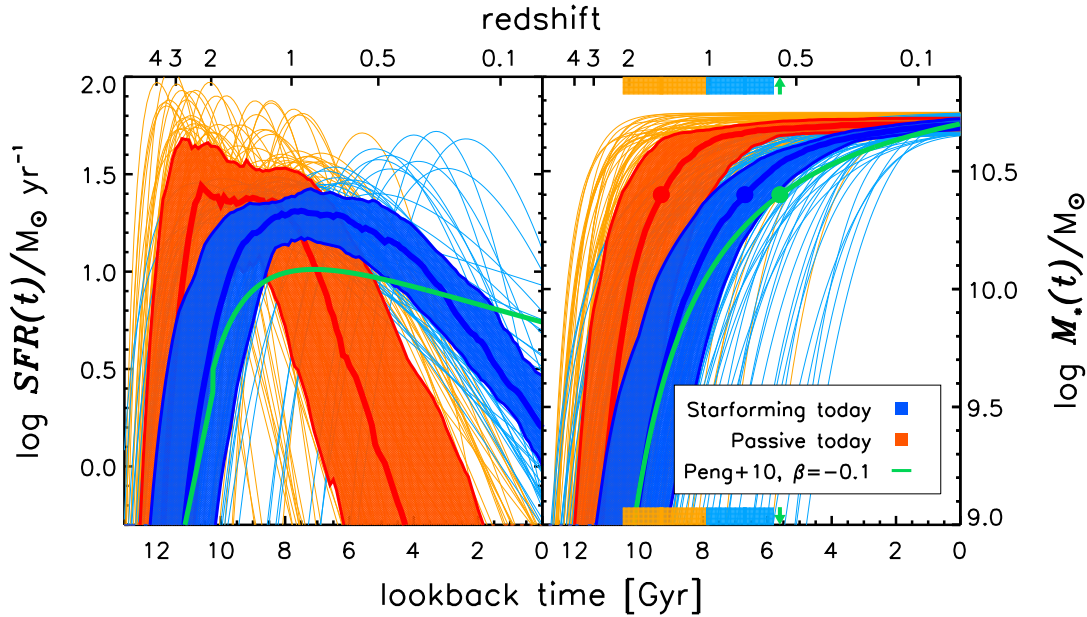}
\caption{A comparison of G13 SFHs ({\it left}) and mass-growth curves ({\it right}) for today's Milky Way-mass galaxies to the result of \citet[][green curve]{PengLilly10}. Turquoise/orange lines denote currently starforming/quiescent G13 galaxies ($\ssfr\geq$ and $< 10^{-11}\,{\rm yr^{-1}}$, \resp); blue and red bands show 25\%/75\% spreads of these SFHs at fixed time. At {\it right}, turquoise/orange bars denote the 25\%/75\% range in half-mass times for the G13 SFHs; green arrows show that for the Peng et al.\ curve. These models paint different astrophysical pictures, but both appear capable of matching important metrics of galaxy evolution. Clearly, a key distinction is the prediction of a wide diversity of SFHs by the galaxy-up G13 model (e.g., a $\sim2$ Gyr inter-quartile range in half-mass times for MW analogs), which is difficult to reproduce from the population-down perspective.}
\label{fig:SFH}
\end{figure*}

\subsection{Fundamentally Different Philosophies}
\label{sec:philo}

To address these questions, we must first understand what distinguishes the aforementioned analyses. Broadly, they fall into two categories: ``population-down'' \citep{PengLilly10,Behroozi13,Lu14} and ``galaxy-up'' \citep[G13;][]{Kelson14}.

The population-down approaches use ensemble averages to infer the behavior of individual systems. They tend to center around a mapping between some ``fundamental'' quantity and $\dot\Mstel(t)$. This mapping can be the \MS\ itself \citep[][]{PengLilly10}, or theoretical dark matter halo growth models combined with, e.g., $\Mhalo$--$\Mstel$ \citep{Behroozi13} relations. Regardless, since they are driven by statistical averages, their ability to assess {\it diversity} in SFHs at fixed $(\Mstel,\sfr,t)$ are limited by construction (Figure \ref{fig:SFH}, green curve). 

By contrast, the galaxy-up approaches \citep[G13;][]{Kelson14} entail no central mappings. They can accommodate (for example) significant/arbitrary scatter in the \MS, and so have great ability to capture diversity in SFHs (Figure \ref{fig:SFH}, turquoise/orange curves; Figure 2 of \citealt{Kelson14}). However, this freedom is paid for by placing demands on the (nature of the) SFHs themselves: \citet{Kelson14} requires them to be governed by quasi-stochastic processes; we implicitly attribute physical importance to the log-normal form. 

The crux of the issue is this: both population-down and galaxy-up approaches have demonstrated significant potential in terms of their ability to match key metrics of galaxy evolution, yet the way they view {\it diversity} -- either in galaxy SFHs or as manifest by dispersion in statistical relations -- are quite different. 

Addressing one question in particular will greatly elucidate either approach's relevance to galaxy evolution:
\begin{quote}
	To what extent did today's equal-mass starforming galaxies ``grow up'' together? 
\end{quote}	
Robust knowledge of the timescale encoded by the \MS\ dispersion, $\sigms$, is critical to addressing this question.

\subsection{What is $\sigms$?}
\label{sec:sigma}

Assuming it does not reflect observational errors,\footnote{Through detailed comparisons of SFRs obtained using different techniques, we are in the process of quantitatively demonstrating this to be true (Oemler et al., in preparation).} there are three interpretations for $\sigms$: (1) short-timescale ($10^{\sim7}$ yr) perturbations, (2) intermediate-timescale ($10^{8-9+}$ yr) variations, or (3) Hubble-timescale ($10^{\sim10}$ yr) differentiation. If (1) is the case, the population-down approach is accurate: today's Milky Way-mass starforming galaxies mostly grew-up together. As such, we may have already identified the most critical astrophysics shaping SFHs \citep{PengLilly10, Behroozi13}. However, if (2) or (3) is the case, the galaxy-up approaches and their consequences must be examined more thoroughly: MW-like galaxies did {\it not} grow-up together and critical astrophysics lies in what diversifies SFHs.

Because galaxy structure (generally) changes on timescales longer than the SFR, one way forward is a detailed morpho-structural study of equal-mass \MS\ galaxies. If $\sigms$ reflects short ``blips'', then there is essentially one way to reach a given $(\Mstel,t)$ and still land on the \MS. Hence, the population-down models suggest there should be little-to-no diversity in, e.g., bulge-to-total mass ratios for these galaxies. In the intermediate-timescale, quasi-stochastic model of \citet{Kelson14}, there are many disparate paths to such an endpoint, suggesting a uniform structural mix. Finally, in the Hubble-timescale case of G13, the paths are numerous, but {\it coherent}. As such, it predicts a trend in structural properties as a function of distance from the mean \MS\ relation.

Many works now point to morphology/structure as as important to shaping the \MS\ \citep{Salmi12,Abramson14a} or pulling galaxies off it \citep{Martig09,Williams10,Wuyts11,Bluck14,Lang14,Omand14}. However, it is still unclear if/how it distinguishes galaxies {\it on} it. Hence, it may relate to ``quenching'' (constant with population-down models) {\it or} ``diversification'' (as implied by galaxy-up models). The goal is to determine which is the case.

\section{Conclusion}

We have shown that a model consisting of individual, log-normal star formation histories entailing no quenching prescription and constrained only at $z\lesssim1$, matches the evolution of the total and quiescent stellar mass functions at $z\lesssim2.5$. It also reproduces the observed SFR--$\Mstel$ relation over the same interval, and is fully consistent with the evolution of the cosmic SFR density $z\leq8$ \citep{Gladders13b}. As such, this achievement makes this ``galaxy-up'' model competitive with other ``population-down'' models, wherein mean ensemble properties describe the evolution of individual systems. We propose that the key difference between the two approaches is the extent to which they stress the importance of diversity in (starforming) galaxy populations. Robust measurements of the size and timescale associated with dispersion in SFR($\Mstel$) could substantially illuminate the extent to which either approach reflects real galaxy SFHs.


\section*{Acknowledgements}
LEA thanks Daniel Kelson for his patience and open-door policy, and acknowledges generous support from the inaugural James Cronin Fellowship.\\


\bibliographystyle{apj}
\small

\begin{thebibliography}{27}
\expandafter\ifx\csname natexlab\endcsname\relax\def\natexlab#1{#1}\fi

\bibitem[{{Abramson} {et~al.}(2014){Abramson}, {Kelson}, {Dressler},
  {Poggianti}, {Gladders}, {Oemler}, \& {Vulcani}}]{Abramson14a}
{Abramson}, L.~E., {Kelson}, D.~D., {Dressler}, A., {et~al.} 2014, \apjl, 785,
  L36

\bibitem[{{Ball} {et~al.}(2006){Ball}, {Loveday}, {Brunner}, {Baldry}, \&
  {Brinkmann}}]{Ball06}
{Ball}, N.~M., {Loveday}, J., {Brunner}, R.~J., {Baldry}, I.~K., \&
  {Brinkmann}, J. 2006, \mnras, 373, 845

\bibitem[{{Behroozi} {et~al.}(2013){Behroozi}, {Wechsler}, \&
  {Conroy}}]{Behroozi13}
{Behroozi}, P.~S., {Wechsler}, R.~H., \& {Conroy}, C. 2013, \apj, 770, 57

\bibitem[{{Bluck} {et~al.}(2014){Bluck}, {Mendel}, {Ellison}, {Moreno},
  {Simard}, {Patton}, \& {Starkenburg}}]{Bluck14}
{Bluck}, A.~F.~L., {Mendel}, J.~T., {Ellison}, S.~L., {et~al.} 2014, \mnras,
  441, 599

\bibitem[{{Brinchmann} {et~al.}(2004){Brinchmann}, {Charlot}, {White},
  {Tremonti}, {Kauffmann}, {Heckman}, \& {Brinkmann}}]{Brinchmann04}
{Brinchmann}, J., {Charlot}, S., {White}, S.~D.~M., {et~al.} 2004, \mnras, 351,
  1151

\bibitem[{{Calvi} {et~al.}(2011){Calvi}, {Poggianti}, \& {Vulcani}}]{Calvi11}
{Calvi}, R., {Poggianti}, B.~M., \& {Vulcani}, B. 2011, \mnras, 416, 727

\bibitem[{{Cucciati} {et~al.}(2012){Cucciati}, {Tresse}, {Ilbert}, {Le
  F{\`e}vre}, {Garilli}, {Le Brun}, {Cassata}, {Franzetti}, {Maccagni},
  {Scodeggio}, {Zucca}, {Zamorani}, {Bardelli}, {Bolzonella}, {Bielby},
  {McCracken}, {Zanichelli}, \& {Vergani}}]{Cucciati12}
{Cucciati}, O., {Tresse}, L., {Ilbert}, O., {et~al.} 2012, \aap, 539, A31

\bibitem[{{Gladders} {et~al.}(2013){Gladders}, {Oemler}, {Dressler},
  {Poggianti}, {Vulcani}, \& {Abramson}}]{Gladders13b}
{Gladders}, M.~D., {Oemler}, A., {Dressler}, A., {et~al.} 2013, \apj, 770, 64

\bibitem[{{Kauffmann} {et~al.}(2003){Kauffmann}, {Heckman}, {White}, {Charlot},
  {Tremonti}, {Brinchmann}, {Bruzual}, {Peng}, {Seibert}, {Bernardi},
  {Blanton}, {Brinkmann}, {Castander}, {Cs{\'a}bai}, {Fukugita}, {Ivezic},
  {Munn}, {Nichol}, {Padmanabhan}, {Thakar}, {Weinberg}, \&
  {York}}]{Kauffmann03}
{Kauffmann}, G., {Heckman}, T.~M., {White}, S.~D.~M., {et~al.} 2003, \mnras,
  341, 33

\bibitem[{{Kelson}(2014)}]{Kelson14}
{Kelson}, D.~D. 2014, ArXiv e-prints

\bibitem[{{Lang} {et~al.}(2014){Lang}, {Wuyts}, {Somerville}, {Forster
  Schreiber}, {Genzel}, {Bell}, {Brammer}, {Dekel}, {Faber}, {Ferguson},
  {Grogin}, {Kocevski}, {Koekemoer}, {Lutz}, {McGrath}, {Momcheva}, {Nelson},
  {Primack}, {Rosario}, {Skelton}, {Tacconi}, {van Dokkum}, \&
  {Whitaker}}]{Lang14}
{Lang}, P., {Wuyts}, S., {Somerville}, R., {et~al.} 2014, ArXiv e-prints

\bibitem[{{Lu} {et~al.}(2014){Lu}, {Mo}, {Lu}, {Katz}, {Weinberg}, {van den
  Bosch}, \& {Yang}}]{Lu14}
{Lu}, Z., {Mo}, H.~J., {Lu}, Y., {et~al.} 2014, ArXiv e-prints

\bibitem[{{Madau} \& {Dickinson}(2014)}]{MadauDickinson14}
{Madau}, P., \& {Dickinson}, M. 2014, \araa, 52, 415

\bibitem[{{Martig} {et~al.}(2009){Martig}, {Bournaud}, {Teyssier}, \&
  {Dekel}}]{Martig09}
{Martig}, M., {Bournaud}, F., {Teyssier}, R., \& {Dekel}, A. 2009, \apj, 707,
  250

\bibitem[{{Moster} {et~al.}(2013){Moster}, {Naab}, \& {White}}]{Moster13}
{Moster}, B.~P., {Naab}, T., \& {White}, S.~D.~M. 2013, \mnras, 428, 3121

\bibitem[{{Moustakas} {et~al.}(2013){Moustakas}, {Coil}, {Aird}, {Blanton},
  {Cool}, {Eisenstein}, {Mendez}, {Wong}, {Zhu}, \& {Arnouts}}]{Moustakas13}
{Moustakas}, J., {Coil}, A.~L., {Aird}, J., {et~al.} 2013, \apj, 767, 50

\bibitem[{{Muzzin} {et~al.}(2013){Muzzin}, {Marchesini}, {Stefanon}, {Franx},
  {McCracken}, {Milvang-Jensen}, {Dunlop}, {Fynbo}, {Brammer}, {Labb{\'e}}, \&
  {van Dokkum}}]{Muzzin13}
{Muzzin}, A., {Marchesini}, D., {Stefanon}, M., {et~al.} 2013, \apj, 777, 18

\bibitem[{{Noeske} {et~al.}(2007){Noeske}, {Weiner}, {Faber}, {Papovich},
  {Koo}, {Somerville}, {Bundy}, {Conselice}, {Newman}, {Schiminovich}, {Le
  Floc'h}, {Coil}, {Rieke}, {Lotz}, {Primack}, {Barmby}, {Cooper}, {Davis},
  {Ellis}, {Fazio}, {Guhathakurta}, {Huang}, {Kassin}, {Martin}, {Phillips},
  {Rich}, {Small}, {Willmer}, \& {Wilson}}]{Noeske07}
{Noeske}, K.~G., {Weiner}, B.~J., {Faber}, S.~M., {et~al.} 2007, \apjl, 660,
  L43

\bibitem[{{Omand} {et~al.}(2014){Omand}, {Balogh}, \& {Poggianti}}]{Omand14}
{Omand}, C.~M.~B., {Balogh}, M.~L., \& {Poggianti}, B.~M. 2014, \mnras

\bibitem[{{Peng} {et~al.}(2010){Peng}, {Lilly}, {Kova{\v c}}, {Bolzonella},
  {Pozzetti}, {Renzini}, {Zamorani}, {Ilbert}, {Knobel}, {Iovino}, {Maier},
  {Cucciati}, {Tasca}, {Carollo}, {Silverman}, {Kampczyk}, {de Ravel},
  {Sanders}, {Scoville}, {Contini}, {Mainieri}, {Scodeggio}, {Kneib}, {Le
  F{\`e}vre}, {Bardelli}, {Bongiorno}, {Caputi}, {Coppa}, {de la Torre},
  {Franzetti}, {Garilli}, {Lamareille}, {Le Borgne}, {Le Brun}, {Mignoli},
  {Perez Montero}, {Pello}, {Ricciardelli}, {Tanaka}, {Tresse}, {Vergani},
  {Welikala}, {Zucca}, {Oesch}, {Abbas}, {Barnes}, {Bordoloi}, {Bottini},
  {Cappi}, {Cassata}, {Cimatti}, {Fumana}, {Hasinger}, {Koekemoer},
  {Leauthaud}, {Maccagni}, {Marinoni}, {McCracken}, {Memeo}, {Meneux}, {Nair},
  {Porciani}, {Presotto}, \& {Scaramella}}]{PengLilly10}
{Peng}, Y.-j., {Lilly}, S.~J., {Kova{\v c}}, K., {et~al.} 2010, \apj, 721, 193

\bibitem[{{Salmi} {et~al.}(2012){Salmi}, {Daddi}, {Elbaz}, {Sargent},
  {Dickinson}, {Renzini}, {Bethermin}, \& {Le Borgne}}]{Salmi12}
{Salmi}, F., {Daddi}, E., {Elbaz}, D., {et~al.} 2012, \apjl, 754, L14

\bibitem[{{Salmon} {et~al.}(2014){Salmon}, {Papovich}, {Finkelstein}, {Tilvi},
  {Finlator}, {Behroozi}, {Dahlen}, {Dav{\'e}}, {Dekel}, {Dickinson},
  {Ferguson}, {Giavalisco}, {Long}, {Lu}, {Reddy}, {Somerville}, \&
  {Wechsler}}]{Salmon14}
{Salmon}, B., {Papovich}, C., {Finkelstein}, S.~L., {et~al.} 2014, ArXiv
  e-prints

\bibitem[{{Tomczak} {et~al.}(2014){Tomczak}, {Quadri}, {Tran}, {Labb{\'e}},
  {Straatman}, {Papovich}, {Glazebrook}, {Allen}, {Brammer}, {Kacprzak},
  {Kawinwanichakij}, {Kelson}, {McCarthy}, {Mehrtens}, {Monson}, {Persson},
  {Spitler}, {Tilvi}, \& {van Dokkum}}]{Tomczak14}
{Tomczak}, A.~R., {Quadri}, R.~F., {Tran}, K.-V.~H., {et~al.} 2014, \apj, 783,
  85

\bibitem[{{Williams} {et~al.}(2009){Williams}, {Quadri}, {Franx}, {van Dokkum},
  \& {Labb{\'e}}}]{Williams09}
{Williams}, R.~J., {Quadri}, R.~F., {Franx}, M., {van Dokkum}, P., \&
  {Labb{\'e}}, I. 2009, \apj, 691, 1879

\bibitem[{{Williams} {et~al.}(2010){Williams}, {Quadri}, {Franx}, {van Dokkum},
  {Toft}, {Kriek}, \& {Labb{\'e}}}]{Williams10}
{Williams}, R.~J., {Quadri}, R.~F., {Franx}, M., {et~al.} 2010, \apj, 713, 738

\bibitem[{{Wuyts} {et~al.}(2011){Wuyts}, {F{\"o}rster Schreiber}, {van der
  Wel}, {Magnelli}, {Guo}, {Genzel}, {Lutz}, {Aussel}, {Barro}, {Berta},
  {Cava}, {Graci{\'a}-Carpio}, {Hathi}, {Huang}, {Kocevski}, {Koekemoer},
  {Lee}, {Le Floc'h}, {McGrath}, {Nordon}, {Popesso}, {Pozzi}, {Riguccini},
  {Rodighiero}, {Saintonge}, \& {Tacconi}}]{Wuyts11}
{Wuyts}, S., {F{\"o}rster Schreiber}, N.~M., {van der Wel}, A., {et~al.} 2011,
  \apj, 742, 96

\bibitem[{{York} {et~al.}(2000){York}, {Adelman}, {Anderson}, {Anderson},
  {Annis}, {Bahcall}, {Bakken}, {Barkhouser}, {Bastian}, {Berman}, {Boroski},
  {Bracker}, {Briegel}, {Briggs}, {Brinkmann}, {Brunner}, {Burles}, {Carey},
  {Carr}, {Castander}, {Chen}, {Colestock}, {Connolly}, {Crocker}, {Csabai},
  {Czarapata}, {Davis}, {Doi}, {Dombeck}, {Eisenstein}, {Ellman}, {Elms},
  {Evans}, {Fan}, {Federwitz}, {Fiscelli}, {Friedman}, {Frieman}, {Fukugita},
  {Gillespie}, {Gunn}, {Gurbani}, {de Haas}, {Haldeman}, {Harris}, {Hayes},
  {Heckman}, {Hennessy}, {Hindsley}, {Holm}, {Holmgren}, {Huang}, {Hull},
  {Husby}, {Ichikawa}, {Ichikawa}, {Ivezi{\'c}}, {Kent}, {Kim}, {Kinney},
  {Klaene}, {Kleinman}, {Kleinman}, {Knapp}, {Korienek}, {Kron}, {Kunszt},
  {Lamb}, {Lee}, {Leger}, {Limmongkol}, {Lindenmeyer}, {Long}, {Loomis},
  {Loveday}, {Lucinio}, {Lupton}, {MacKinnon}, {Mannery}, {Mantsch}, {Margon},
  {McGehee}, {McKay}, {Meiksin}, {Merelli}, {Monet}, {Munn}, {Narayanan},
  {Nash}, {Neilsen}, {Neswold}, {Newberg}, {Nichol}, {Nicinski}, {Nonino},
  {Okada}, {Okamura}, {Ostriker}, {Owen}, {Pauls}, {Peoples}, {Peterson},
  {Petravick}, {Pier}, {Pope}, {Pordes}, {Prosapio}, {Rechenmacher}, {Quinn},
  {Richards}, {Richmond}, {Rivetta}, {Rockosi}, {Ruthmansdorfer}, {Sandford},
  {Schlegel}, {Schneider}, {Sekiguchi}, {Sergey}, {Shimasaku}, {Siegmund},
  {Smee}, {Smith}, {Snedden}, {Stone}, {Stoughton}, {Strauss}, {Stubbs},
  {SubbaRao}, {Szalay}, {Szapudi}, {Szokoly}, {Thakar}, {Tremonti}, {Tucker},
  {Uomoto}, {Vanden Berk}, {Vogeley}, {Waddell}, {Wang}, {Watanabe},
  {Weinberg}, {Yanny}, {Yasuda}, \& {SDSS Collaboration}}]{York00}
{York}, D.~G., {Adelman}, J., {Anderson}, Jr., J.~E., {et~al.} 2000, \aj, 120,
  1579

\end{thebibliography}


\clearpage
\end{document}